# Impact of Spherical Coordinates Transformation Pre-processing in Deep Convolution Neural Networks for Brain Tumor Segmentation and Survival Prediction

Carlo Russo[1*][0000-0001-8296-4345], Sidong Liu[1,2*][0000-0002-2371-0713], Antonio Di Ieva [1][0000-0002-5341-5416]

[1] Computational NeuroSurgery (CNS) Lab, Macquarie University, Sydney, Australia
[2] Centre for Health Informatics, Australian Institute of Health Innovation, Macquarie University, Sydney, Australia

**Abstract.** Pre-processing and Data Augmentation play an important role in Deep Convolutional Neural Networks (DCNN). Whereby several methods aim for standardization and augmentation of the dataset, we here propose a novel method aimed to feed DCNN with spherical space transformed input data that could better facilitate feature learning compared to standard Cartesian space images and volumes. In this work, the spherical coordinates transformation has been applied as a preprocessing method that, used in conjunction with normal MRI volumes, improves the accuracy of brain tumor segmentation and patient overall survival (OS) prediction on Brain Tumor Segmentation (BraTS) Challenge 2020 dataset. The LesionEncoder framework has been then applied to automatically extract features from DCNN models, achieving 0.586 accuracy of OS prediction on the validation data set, which is one of the best results according to BraTS 2020 leaderboard.

## 1 Introduction

Magnetic Resonance Imaging (MRI) is used in everyday clinical practice to assess brain tumors. However, the manual segmentation of each volume representing the extension of the tumor is time-demanding and operator-dependent, as it is often non-reproducible and depends upon neuroradiologists' expertise. Several automatic or semi-automatic segmentation algorithms have been introduced to help segment brain tumors, and Deep Convolutional Neural Networks (DCNN) have recently shown very promising results. To further improve the accuracy of automatic methods, the Multimodal Brain Tumor Segmentation (BraTS) challenge [1-3] is organized annually within the International Conference on Medical Image Computing and Computer Assisted Intervention (MICCAI). The BraTS 2020 challenge includes a task for automatic segmentation of the total area containing the tumor (Whole Tumor - WT), as well as the Necrosis and

---

* Authors contributed equally.

Active tumor cells area (Tumor Core - TC, Enhancing Tumor - ET, Necrosis and ET are contained in TC).

Furthermore, glioma patients often have a dire survival prognosis following surgical resection and radiochemotherapy [4]. Thus, a further task to predict patient overall survival (OS) has been added into the challenge, aimed at improving the prediction of patient survival outcome in order to add information that are relevant to the decision-making process.

DCNNs are data driven algorithms. They require huge amount of data to obtain good results. In medical imaging, such big datasets are not often available, thus pre-processing and data augmentation plays an important role. While pre-processing methods are usually used to standardize input data, they can also be used to enhance meaningful data inside the original input images: an example is cropping the region of interest when the input data includes lots of redundant and misleading information.

Therefore, we propose a novel spherical space transformation method to enhance information on specific points of the tumor as well as enable the DCNN learning process to be invariant to rotation and scaling of the input images. Furthermore, we extended the use of lesion features extracted from the latent space of the segmentation models using the LesionEncoder framework, which replaces the classic imaging / radiomic features, such as volumetric parameters, intensity, morphologic, histogram-based and textural features, which showed high predictive power in patient OS prediction.

## 2 Brain Tumor Segmentation

### 2.1 Background

**Dataset** The dataset consists of four MRI sequences used to determine the segmentation and extract survival features, namely T1-weighted, post-contrast T1, T2-weighted and FLAIR images. Training dataset have 336 4-channel volumes with ground truth segmentation. Validation dataset is composed by data from 125 patients [5,6]. Testing dataset is composed by additional 166 patients.

**Baseline Model** The DCNN that we chose to use as baseline for our method is derived from Myronenko [7], which is based on Variational Auto Encoder (VAE) U-Net with adjusted input shape and loss function according to the type of transformation used into the preprocessing phase. The VAE proposed by Myronenko is composed by a U-Net with two decoder branches: a segmentation decoder branch, used to obtain the final segmentation, and an additional decoder branch to reconstruct the original volumes, used to regularize the shared encoder. The loss function is given by the formula:

$$L = L_{dice} + 0.1 * L_{L2} + 0.1 * L_{KL}$$

where $L_{L2}$ is the L2 loss on the VAE branch and $L_{KL}$ is the KL divergence penalty term.

**Trained Models** We trained different models by changing pre-processing method (Cartesian and spherical) and some layers hyperparameters. Although the models share the same VAE structure proposed by Myronenko, there are a few differences. More specifically, Cartesian_v1 includes standard Dropout with rate 0.2, Kernel size filters 3x3x3 in Convolution layers outside each ResNet-like block of the VAE and an additional 3x3x3 convolution layer before the final output block. While Cartesian_v2 uses SpatialDropout3D with the same ratio, 1x1x1 convolution filters and no additional layers. The spherical model has the same structure of Cartesian_v1 but with Spherical transformation pre-processing on inputs. The spherical transformed model (using Cartesian_v2 structure) has not been trained yet by the 2020 challenge deadline. A bisCartesian model has been also trained using the structure of the Cartesian v1 model but using a lower coefficient of the KL loss, set to 0.0001. The model has not given better segmentation results, although it is showing improved results on OS task.

## 2.2 Spherical Coordinates Transformation

Our team previously presented the spherical coordinate transformation pre-processing as a method to improve segmentation results [8]. The Spherical transformed volume is shown in **Figure 1**.

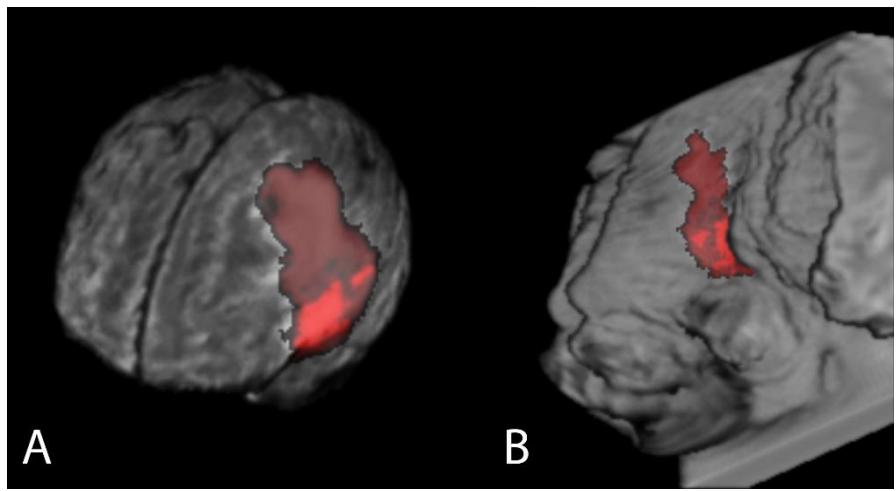

**Figure 1**. Example of the representation of a radiologic volume in spherical coordinate system. A)) A brain MRI volume with its 3D segmentation of the tumor, and B) the same volume transformed into a spherical coordinate system using the center of the volume as the origin.

Each pre-processed volume uses an origin point. Thus, to achieve good performance on the training, it is important to correctly select origin points, included within the tumor. For this reason, we used a cascade of three DCNNs, the first one predicting a coarse segmentation, and then refining the segmentation by using origin points included in the previous model. The first pass model of the cascade could also be a model trained on non-transformed input (a Cartesian model), but the use of the Spherical model already in the first cascade's pass enabled pre-training weights to be used for the next training steps.

The Spherical coordinate transformation also adds an extreme augmentation. This is a beneficial step as it adds invariance to the rotation and scaling to the DCNN model. However, such an invariance also has a drawback especially when dealing with WT segmentation: apparently, WT segmentation works better with a Cartesian model, whereas using Spherical pre-processing adds many false positive regions to the WT. Thus, we used a Cartesian model to filter out the false positive regions found by the spherical pre-processing as shown in **Figure 2**.

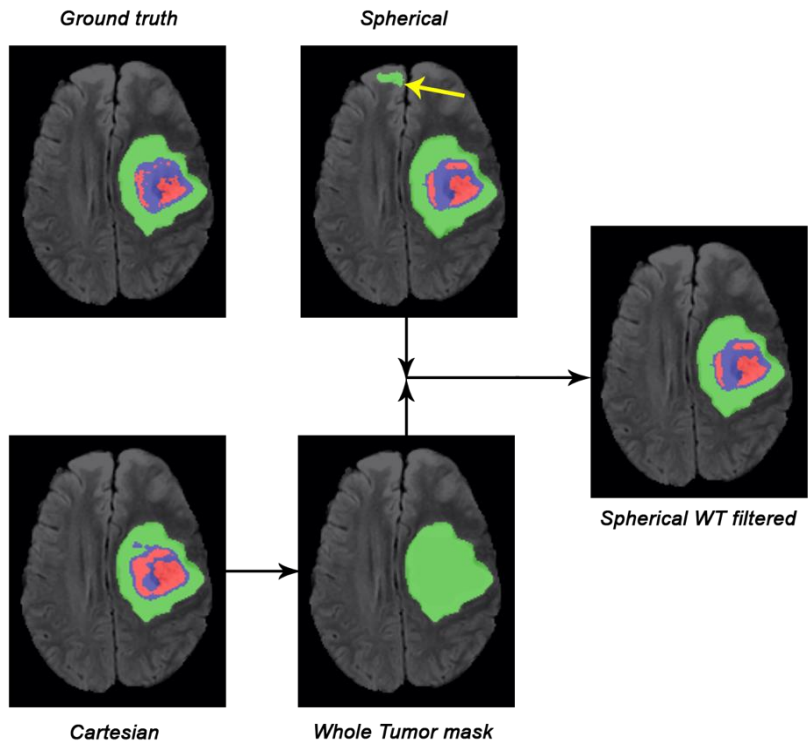

**Figure 2**. An example of the application of the Cartesian filter to the spherical model's segmentation to obtain final segmentation. The final result is obtained by using the Cartesian CNN to filter the WT mask obtained on spherical segmentation. Such a step helps to erase false positive voxel regions wrongly selected by the spherical CNNs (see segmented object indicated by the yellow arrow).

We used this proposed method for the first time in a BraTS challenge, achieving similar results obtained in our original paper regarding the improvement of the accuracy of the model trained on transformed input compared to the baseline model [8]. We also tested the intersection of the segmentation on the three different classes (Spherical - Cartesian intersection 3CH) instead of filtering only the WT class. Finally, we ensembled the best segmentations from Cartesian_v2 and the intersection method to improve the results further.

## 2.3    Post-processing

After filtering the Spherical segmentation with the Cartesian filter, we used a post-processing method to improve ET segmentation. We noticed that many false positive ET segmentations are due to isolated voxels. For this reason, we applied a binary opening operator to isolate thin branches over ET spots and then filter out the spots having less than 30 voxels. When the ET segmentation is still present after these filters, the original ET segmentation is restored and used as the final one. Otherwise, the ET segmentation is completely erased, meaning that no ET is present in the current volume.

## 2.4    Segmentation Results

**Table 1** shows the summary of segmentation results on the validation dataset. The most promising methods tested so far were the Cartesian_v2 and the spherical models: used alone without post-processing, the Cartesian_v2 gave the best results in WT and TC segmentation, while the spherical model worked better on ET segmentation.

Merging Spherical model labels with Cartesian model WT label filter gave a further improvement in the segmentation of ET class, while the best result for the WT and TC classes segmentation remained using only the Cartesian_v2 model.

The best overall improvement on ET has been shown by post-processing the segmentation of the intersected Cartesian and Spherical models on the three channels (3CH), even if the TC class dice score decreased and WT class did not improve further. The final model used on the testing dataset is an Ensemble of methods taking the ET label from the 3CH model and WT and TC labels from the Cartesian v2 model.

**Table 1**. Evaluation of segmentation performance on BraTS 2020 official validation dataset

| Model names | Dice | | | Sensitivity | | | Specificity | | | Hausdorff95_ET | | |
|---|---|---|---|---|---|---|---|---|---|---|---|---|
| | ET | WT | TC | ET | WT | TC | ET | WT | TC | ET | WT | TC |
| Cartesian only (v1) | 0.7046 | 0.8904 | 0.7852 | 0.7204 | 0.9193 | 0.7829 | 0.9997 | 0.9988 | 0.9996 | 39.8783 | 7.6745 | 15.8352 |
| Cartesian only (v2) | 0.7072 | **0.8999** | **0.8250** | 0.7187 | 0.8947 | 0.8324 | 0.9997 | 0.9992 | 0.9995 | 46.4816 | 6.0702 | 8.1284 |
| Cartesian only (v1 bis) | 0.6926 | 0.8954 | 0.7905 | 0.6953 | 0.9064 | 0.7860 | 0.9997 | 0.9991 | 0.9996 | 44.5790 | 6.7375 | 8.2056 |
| Spherical only | 0.7428 | 0.8555 | 0.7783 | **0.7882** | **0.9419** | **0.8455** | 0.9995 | 0.9975 | 0.9991 | 23.3944 | 9.6847 | 8.4968 |
| Spherical with Cartesian v2 WT filter | 0.7533 | 0.8983 | 0.7846 | 0.7867 | 0.8807 | 0.8302 | 0.9995 | **0.9993** | 0.9992 | **22.4573** | **5.9040** | 8.1578 |
| Cartesian v2 with postprocessing | 0.7604 | **0.8999** | **0.8250** | 0.7747 | 0.8947 | 0.8324 | 0.9997 | 0.9992 | 0.9995 | 33.8563 | 6.0702 | 8.1284 |
| Spherical -Cartesian intersection 3CH and postprocessing | **0.7662** | 0.8983 | 0.8060 | 0.7471 | 0.8807 | 0.7803 | **0.9998** | **0.9993** | **0.9997** | 27.3789 | **5.9040** | **7.2708** |
| Ensemble of methods | **0.7662** | **0.8999** | **0.8250** | 0.7471 | 0.8947 | 0.8324 | **0.9998** | 0.9992 | 0.9995 | 27.3789 | 6.0702 | 8.1284 |

Results of the final model on testing dataset shown in **Table 2** seems to confirm a good accuracy, above all on the ET segmentation, although is not possible to make a comparison with the other models since the Challenge only allows to test one method on the dataset.

**Table 2**. Evaluation of segmentation performance on BraTS 2020 official testing dataset

| | Dice | | | Sensitivity | | | Specificity | | | Hausdorff95_ET | | |
|---|---|---|---|---|---|---|---|---|---|---|---|---|
| | ET | WT | TC | ET | WT | TC | ET | WT | TC | ET | WT | TC |
| Mean | 0.7898 | 0.8687 | 0.8066 | 0.7864 | 0.8970 | 0.8515 | 0.9997 | 0.9990 | 0.9995 | 17.9747 | 6.7349 | 22.2474 |
| StdDev | 0.1951 | 0.1373 | 0.2609 | 0.2095 | 0.1283 | 0.2291 | 0.0003 | 0.0011 | 0.0012 | 74.7789 | 10.4801 | 74.6658 |
| Median | 0.8338 | 0.9128 | 0.9068 | 0.8480 | 0.9322 | 0.9285 | 0.9998 | 0.9993 | 0.9997 | 1.7321 | 3.6056 | 3.0000 |
| 25quantile | 0.7508 | 0.8581 | 0.8310 | 0.7442 | 0.8825 | 0.8438 | 0.9996 | 0.9986 | 0.9995 | 1.4142 | 2.0590 | 1.4937 |
| 75quantile | 0.8897 | 0.9412 | 0.9463 | 0.9051 | 0.9677 | 0.9748 | 1.0000 | 0.9996 | 0.9999 | 3.0000 | 6.5312 | 7.2111 |

## 3 Prediction of Patient OS

Our team also participated in Task 2 of the BraTS Challenge: prediction of patient overall OS from pre-operative MRI scans. Instead of using the pre-defined imaging / radiomic features, such as volumetric parameters, intensity, morphologic, histogram-based and textural features, we used the features automatically extracted from MRI scans using the novel LesionEncoder (LE) framework [9]. The LE features were further processed using Principle Component Analysis (PCA) to reduce dimensionality, and then used as input to a generalized linear model (GLM) [10] to predict patient OS.

### 3.1 LesionEncoder Framework

The LE framework was proposed in a recent work for COVID-19 severity assessment and progression prediction [9]. The original LE adopted the U-Net structure [11], which consists of an encoder and a decoder based on the EfficientNet [12]. While the encoder learns and captures the lesion features in the input images, the decoder maps the lesion features back to the original image space and generates the segmentation maps. The features learnt by the encoder in the latent space encapsulate rich information of the lesions, therefore, can be used for lesion segmentation, as well as other tasks such as classification and prediction.

In this study, we used the VAE as backbone to build the LE. As described in the previous section, three different configurations have been applied to the VAE model, resulting in three different lesion encoders: i.e., LE_Cartesian (Cartesian_v2), LE_Spherical (Spherical) and LE_bisCartesian (Cartesian_v1_bis). The latent variables of the input images / MRI scans extracted by individual lesion encoders were then used as the features to predict patient OS. For each MRI scan, a high-dimensional feature vector ($d = 256$) was derived. As the high-dimensional feature space tended to lead to overfitting, we therefore used PCA to control the feature dimensionality by setting different numbers of principle components ($\hat{d} = [2, 60]$ in this study) for further analysis.

## 3.2 Tweedie Regressor

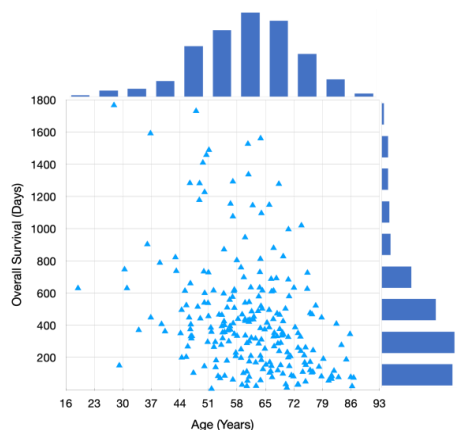

**Figure 3** shows the joint age and OS distribution of the patients in the training cohort. The age distribution, as shown at the top of the figure, seems to be a normal distribution. The OS distribution, on the right side of the figure, seems to be heavily skewed, with the majority of cases having OS less than 400 days. To model the tailed distribution of the OS values, we therefore used a Tweedie distribution [13], a special case of exponential dispersion models whose skewness can be controlled by a power parameter ($r = [1.1, 1.9]$ in this study).

**Figure 3**. Joint age and OS distribution of the training samples.

A GLM model [10] based on the Tweedie distribution, i.e., Tweedie Regressor [13], was built to predict OS values. The Tweedie Regressor was implemented using scikit-learn (v0.23.2). As the resection status and age are essential predictors of OS, both of them were merged with the LesionEncoder features as input to the Tweedie Regressor for OS prediction.

## 3.3 Performance Evaluation

Two evaluation schemes were used to assess the prediction performance. The results were first evaluated based on accuracy of the classification of subjects as long-survivors (>15 months / 450 days), short-survivors (<10 months / 300 days), and mid-survivors (survival rate between 10 and 15 months / 300 – 450 days). In addition, a pairwise error analysis between the predicted and actual OS (in days) was performed, evaluated using the following metrics: mean square error (MSE), median square error (median SE), standard deviation of the square errors (std SE), and the Spearman correlation coefficient (Spearman R). Amongst the 235 patients in the training set, 118 underwent a surgical gross total resection (GTR) and 10 underwent a subtotal resection (STR); in 107 cases, no information about the resection status are available. All of the 29 subjects in the validation set had a GTR resection status. The extent of resection was considered in the model as it has been shown to correlate to post-surgical outcome [14].

### 3.4 OS Prediction Results

**Cross-Validation on the Training Set** We used 5-fold cross-validation to train and validate the proposed method. An internal validation set (20%) was split from the dataset in each fold with the remaining 80% as the training set. For each of the three lesion encoders, i.e., $LE_{Cartesian}$, $LE_{Spherical}$ and $LE_{bisCartesian}$, this process was repeated 5 times, leading to 5 different sub-models. **Figure 4** illustrates the projected feature space of the features extracted using $LE_{Spherical}$ (a), and the scatter plots (b, c) of the predicted OS vs. actual OS of the training samples. There was high variance in the performance of the sub-models (from 0.362 to 0.574).

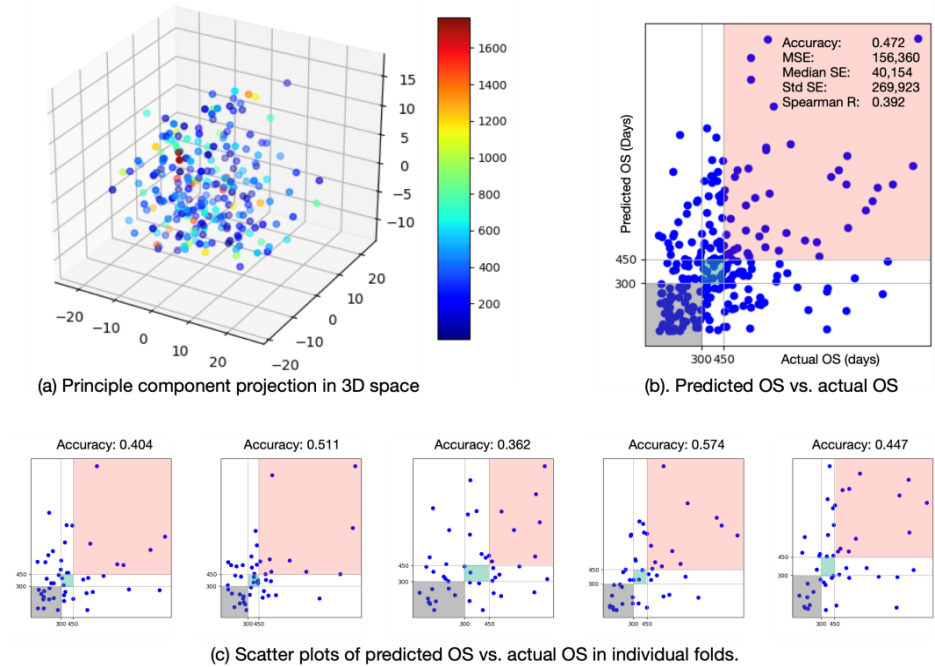

(a) Principle component projection in 3D space

(b). Predicted OS vs. actual OS

(c) Scatter plots of predicted OS vs. actual OS in individual folds.

**Figure 4.** Five-fold cross-validation on the training set using the lesion encoder with spherical configuration ($LE_{Spherical}$). (a) 3D projection of the top 3 principle components derived from PCA. The color table indicates the patients' OS (days). (b) Scatter plot of the predicted OS vs. actual OS of the aggregate of holdout samples in individual folds. Gray, green and red boxes

indicate the correct predictions of the short-, mid- and long-survivors, respectively. (c) Scatter plots of the holdout samples in each cross-validation fold.

The prediction results of the 5 sub-models were further aggregated, and the results were summarized in **Table 3**. LE$_{Cartesian}$ achieved the highest accuracy (0.494) and Spearman R (0.429), while LE$_{Spherical}$ had the lowest MSE, median SE and std SE. These two models outperformed LE$_{bisCartesian}$; however, the differences were not substantial (<0.034 in accuracy).

**Table 3**. Summaries of OS prediction performance on the training set.

| Algorithm | Accuracy | MSE | median SE | std SE | Spearman R |
|---|---|---|---|---|---|
| LE$_{Cartesian}$ | **0.494** | 164,806.75 | 43,224.50 | 402,510.04 | **0.429** |
| LE$_{Spherical}$ | 0.472 | **156,359.53** | **40,153.61** | **269,922.79** | 0.392 |
| LE$_{bisCartesian}$ | 0.460 | 174,067.83 | 49,956.34 | 318,133.99 | 0.329 |

**Prediction Performance on the Validation Set** The 5 different sub-models with the same configuration were then applied to the official validation set (n=29) to predict the OS of each validation case. The mean value of the 5 predictions of each case were then averaged to derive the final prediction. Results of the three models with different configurations are summarized in **Table 4**. The LE$_{bisCartesian}$ model achieved the highest accuracy (0.552); however, its MSE and std SE were higher, and Spearman R was lower than the other models. The LE$_{Cartesian}$ model had the lowest MSE and the highest Spearman R, showing a better representation of the overall distribution of the OS values.

These findings showed a complementary nature of different models; therefore, we combined the outputs of these models to test whether the prediction performance could be improved further. Four combinations were tested, which consistently showed equal or better accuracy (between 0.552 and 0.586) compared to using individual models.

**Table 4**. Summaries of OS prediction results of the validation dataset.

| Algorithm | Accuracy | MSE | median SE | std SE | Spearman R |
|---|---|---|---|---|---|
| LE$_{Cartesian}$ (M1) | 0.483 | **75,908.78** | 32,322.27 | **90,074.92** | **0.498** |
| LE$_{Spherical}$ (M2) | 0.517 | 125,308.38 | **19,226.91** | 216,570.80 | 0.374 |
| LE$_{bisCartesian}$ (M3) | 0.552 | 130,941.15 | 27,762.22 | 232,271.86 | 0.261 |
| M1 & M2 | **0.586** | 88,311.58 | 27,114.54 | 142,969.74 | 0.412 |
| M1 & M3 | **0.586** | 93,565.43 | 24,579.67 | 145,669.08 | 0.360 |
| M2 & M3 | 0.552 | 125,659.88 | 32,131.13 | 223,333.91 | 0.300 |
| M1 & M2 & M3 | **0.586** | 99,776.58 | 24,539.13 | 168,419.90 | 0.338 |

Our final submission for OS prediction on the validation dataset, which was based on the **M1&M2** model, was ranked the 4[th] place in accuracy among the 42 participating teams. In the meanwhile, it achieved the 5[th] place in both MSE and Spearman R, the 8[th] place in median SE, and the 10[th] place in std SE (checked on 23 October 2020).

**Table 5**. Summaries of OS prediction results of the testing dataset.

| Algorithm | Accuracy | MSE | median SE | std SE | Spearman R |
|---|---|---|---|---|---|
| M1 & M2 & M3 | 0.495 | 469,253.36 | 67,626.86 | 1,281,569.74 | 0.379 |

We further applied the **M1&M2&M3** model on the official test dataset (n=107). The model's performance, as shown in **Table 5**, was lower on the test dataset compared to the validation dataset, implying a marked difference between the two datasets and overfitting of the model. However, without knowing the results of other models, either of our own or from other participating teams, it is difficult to confirm whether such performance drop is caused by a less representative training dataset or a less generalizable model, or both.

## 4    Discussion

**Spherical Coordinate Transformation Pre-processing** Spherical coordinate transformation pre-processing of the input dataset contribute to explore data in a different way, thus changing the learning process and achieving different features compared to the classical DCNN model learning process. Those different features can help to improve the segmentation process as well as contributing to deep feature extraction to be used in patients' OS prediction. Even if the spherical pre-processing method contributes to improving baseline model results, simple post-processing methods also have a strong impact on segmentation accuracy. However, overall segmentation results obtained by this method are not amongst the best ones compared to other teams in BraTS 2020 leaderboard, and additional efforts should be done to fine tune both Cartesian and spherical training phase.

**LesionEncoder Framework** The LesionEncoder framework extends the use of lesion features beyond conventional lesion segmentation. There is a wealth of information in the brain tumors including shape, texture, location, extent and distribution of involvement of the abnormality, that can be extracted by the lesion encoder. While it has been demonstrated in COVID-19 progression prediction [9] and severity assessment [15], here we demonstrated a new application of LE in patient OS prediction. It may have strong potential in a wide range of other clinical and research applications, e.g., brain tumor pseudo-progression detection [16] and ophthalmic disease screening [17].

**Model Optimization** Various dimension reduction methods have been tested in this study, including PCA, Independent Component Analysis (ICA), t-distributed stochastic neighbor embedding (T-SNE). In the training phase, PCA was found to have lower variability in accuracy than the other methods; as a result, it was chosen to process the high-dimensional features. We used a linear search strategy to optimize the two most important parameters of the OS prediction model, including the number of principle components in PCA ($\hat{d} = [2, 60]$) and the power of Tweedie distribution ($r = [1.1, 1.9]$). The optimal parameters for LE$_{\text{Cartesian}}$ were ($\hat{d} = 10, r = 1.6$), and ($\hat{d} = 3, r = 1.6$) for both LE$_{\text{Spherical}}$ and LE$_{\text{bisCartesian}}$. In addition, it will be important to demonstrate the scale invariance in the Tweedie regressor within different datasets in our future work.

# 5    Conclusion

In conclusion, we have introduced a novel and very promising method to pre-process brain tumors' MR images by means of a spherical coordinates transformation to be used in DCNN models for brain tumor segmentation. The LesionEncoder framework has been applied to automatically extract imaging features from DCNN models, demonstrating good performance for the survival prediction task.